\documentclass[aps,prb,twocolumn,showpacs]{revtex4}

\usepackage{graphicx}

\def\vR{{\bf R}}
\def\vS{{\bf S}}
\def\vn{{\bf n}}

\begin{document}

\title{Spin-orientation-dependent spatial structure of a magnetic acceptor state in a zincblende semiconductor}

\author{Jian-Ming Tang and Michael E. Flatt\'e}

\affiliation{Department of Physics and Astronomy, University of Iowa,
  Iowa City, Iowa 52242-1479, USA}

%\date{\today}

\begin{abstract}

The spin orientation of a magnetic dopant in a zincblende
semiconductor strongly influences the spatial structure of an acceptor
state bound to the dopant. The acceptor state has a roughly oblate
shape with the short axis aligned with the dopant's core spin. For a
Mn dopant in GaAs the local density of states at a site $8$~\AA{} away
from the dopant can change by as much by $90$\% when the Mn spin
orientation changes. These changes in the local density of states
could be probed by scanning tunneling microscopy to infer the magnetic
dopant's spin orientation.

\end{abstract}


\maketitle

Semiconductors doped with magnetic atoms have attracted much interest
in recent years because of their potential applications in spintronic
and quantum information technology.\cite{Wolf2001,Awschalom2002} The
magnetic dopants introduce, in addition to magnetic moments,
spin-polarized acceptor states into the semiconductor
host.\cite{Vogl1985,Zunger1986,Schneider1987,Linnarsson1997,Mahadevan2004a}
Recent advances in scanning tunneling microscopy (STM) on GaAs have
led to a better understanding of the anisotropic shape of these
acceptor
states.\cite{Yakunin2004a,Yakunin2004b,Kitchen2005,Yakunin2005} The
principal determinant of the anisotropic shape is the cubic symmetry
of the lattice, and not the spin-orbit interaction, as the
characteristic anisotropic shape is predicted to form even in
semiconductors with negligible spin-orbit
interaction.\cite{Yakunin2004b} However, the spin-orbit interaction
does partly correlate the degree of acceptor state anisotropy with the
spin orientation of the magnetic dopant.\cite{JMT2004} This
correlation suggests the possibility of detecting the spin orientation
of a magnetic dopant with a purely nonmagnetic probe.  A similar
phenomenon has been found in metallic magnetic systems. The spin-orbit
interaction in iron films mixes the $d$ bands to yield a few percent
difference in the LDOS for nanoscale magnetic domains (out-of-plane
magnetization) and domain walls (in-plane
magnetization).\cite{Bode2002} A coupling between spin and orbital
degrees of freedom in quantum dots has been predicted to generate a
spin-dependent electric field, although the size of that effect has
been estimated to be very small.\cite{Levitov2003} A
spin-orientation-dependent LDOS has yet to be observed in
semiconductor systems.

Here we describe calculations of the dependence of the LDOS near a Mn
dopant in GaAs on the Mn spin orientation. The orientation-dependence
of the LDOS ranges from very high ($\sim 90$\%) for tunneling into the
acceptor state, to quite small ($\sim <5$\%) for tunneling into
continuum valence states far from the band edge.  Tunneling into the
continuum valence states resembles the qualitative behavior seen in
iron films;\cite{Bode2002} the LDOS orientation-dependence is largest
near critical points of the electronic structure: near the valence
band edge for GaAs:Mn, and near an avoided level crossing for
iron. Tunneling into discrete states such as the acceptor level,
however, can be highly selective of spin and orbital character, as
localized states with different orbital characters have narrow enough
linewidths not to be spectrally distinguishable. Electron spin resonance
(ESR) of the Mn spins has been observed\cite{Schneider1987} in GaMnAs;
thus we predict that the ESR of a single Mn spin could be detected using a
nonmagnetic scanning tunneling microscope.

\begin{figure}
\includegraphics[width=\columnwidth]{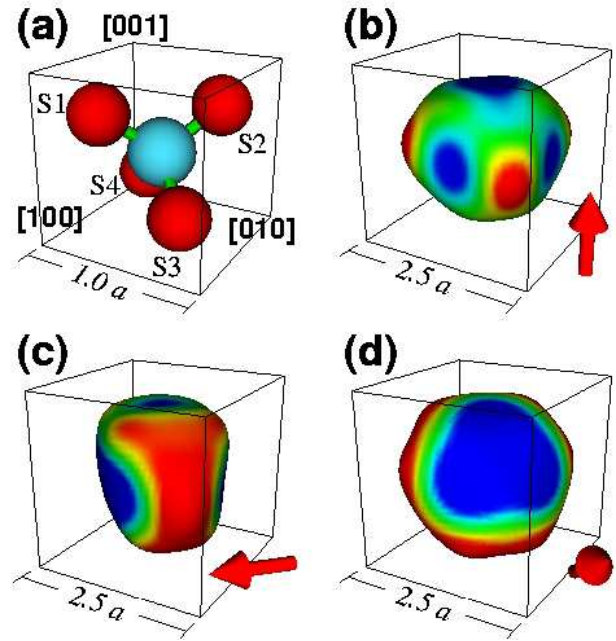}
\caption{ (a) Atomic structure near a substitutional Mn dopant (blue)
in the GaAs lattice (red atoms are As). The As atoms are labeled by
S1, S2, S3, and S4. (b--d) Contour surfaces of the LDOS of the
acceptor level at 10\% of the peak value at the Mn site.  The Mn spin
is aligned with the (b) $[001]$, (c) $[1\bar{1}0]$ or (d) $[111]$ axis
of the GaAs lattice. The symmetry is (b) $D_{2d}$, (c) $C_{2v}$ or (d)
$C_{3v}$. The LDOS at each atomic site is spatially distributed
according to a normalized Gaussian with a $2.5$\AA{} width. The box
outlines are aligned with the cubic lattice and have widths in units
of the lattice constant ($a=5.65$\AA). }
\label{fig:Mn3D}
\end{figure}

The results presented in this paper are calculated from a $sp^3$
tight-binding model, including on-site spin-orbit coupling, for an
isolated Mn acceptor state in GaAs.\cite{JMT2004} The Mn spin is
assumed to be a classical spin aligned in a certain direction, for
example by applying an external magnetic field. The Mn spin
degree-of-freedom is treated as an effective spin-dependent potential
acting on the valence electrons of GaAs. Motivated by a $p$-$d$
hybridization model,\cite{Vogl1985} our Mn potential has
spin-dependent matrix elements at the four nearest-neighbor sites. The
Hamiltonian with a single Mn dopant takes the form
\begin{eqnarray}
H & = & H_0 + V_{\hat\vS} \;,
\end{eqnarray}
where $H_0$ is the $sp^3$ tight-binding Hamiltonian for bulk GaAs with
only nearest-neighbor hopping,\cite{Chadi1977} and $V_{\hat\vS}$ is
the effective potential due to the Mn dopant with its spin pointing in
the direction $\hat\vS$.  When the Mn spin is aligned with the $[001]$
crystal axis, the potential is
\begin{eqnarray}
V_{[001]} & = & V_n\otimes{\bf 1}+V_m\otimes\sigma_3 \;,
\end{eqnarray}
where $V_n$ and $V_m$ are the spatial part (including both lattice
sites and atomic orbitals) of the nonmagnetic and magnetic potential,
and $\sigma_j$'s are Pauli spin matrices. In our model, $V_n$ is
non-zero at the Mn and the four nearest-neighbor As sites, and $V_m$
is only non-zero at the four nearest-neighbor As sites. The non-zero
potential matrix elements at the neighboring As sites come from the
hybridization among the Mn $d$ orbitals and the $sp^3$ hybrids of
GaAs. The potential $V_{\hat\vS}$ for the Mn spin pointing in a
general direction can then be obtained through a rotation in the spin
space,
\begin{eqnarray}
V_{\hat\vS} & = & U_{\hat\vS}V_{[001]}U_{\hat\vS}^\dagger \;, \label{eq:pot}
\end{eqnarray}
where
\begin{eqnarray}
U_{\hat\vS} & = & \exp\left(-i\frac{\vec\sigma}{2}\cdot\hat\vn\theta\right) \;,
\end{eqnarray}
$\theta$ is the angle between the vector $\hat\vS$ and the $[001]$
axis, and $\hat\vn$ is a unit vector pointing to the direction of
$[001]\times\hat\vS$.  We find that adding the Coulomb potential out
to the 5th nearest neighbor to the impurity (and correspondingly
reducing $V_m$ to keep the acceptor binding energy fixed) does not
significantly alter the anisotropy of the acceptor state we describe
in this paper and that is visible in Figs. 1-4.

Within this $sp^3$ model, the energy spectrum of the system is
independent of the Mn spin orientation. This is a direct consequence
of the ability to transform two Hamiltonians with different Mn spin
orientations into each other with a unitary transformation, which we
now derive. The basis states ($s$ and $p$ orbitals) are irreducible
representations of the $T_d$ group and of the rotation group SO(3). As
a result, the Mn potential is invariant under a rotation in real
space,
\begin{eqnarray}
V_{\hat\vS} & = & U_R^\dagger V_{\hat\vS} U_R \;, \label{eq:UR}
\end{eqnarray}
where $U_R = U^L_{\vR^\prime \vR}\otimes U^o_{l^\prime l}$. The
unitary transformations $U^L$ and $U^o$ correspond to the rotations of
the lattice and of the atomic orbitals respectively. Note that the
homogeneous Hamiltonian $H_0$ including the spin-orbit interaction is
invariant under a full rotation,
\begin{eqnarray}
H_0 & = & U^\dagger H_0 U \;,
\end{eqnarray}
where
\begin{eqnarray}
U & = & U^L_{\vR^\prime,\vR}\otimes U^o_{l^\prime,l}\otimes U^s_{s^\prime,s} \;.
\end{eqnarray}
With Eqs.~(\ref{eq:pot}) and (\ref{eq:UR}), the rotation of the Mn
spin can be achieved via rotating the whole lattice in the opposite
direction,
\begin{eqnarray}
H_0 + V_{\hat\vS} & = & U^\dagger(H_0 + V_{[001]})U \;,
\end{eqnarray}
if $U^s = U^\dagger_{\hat\vS}$.

\begin{figure}
\includegraphics[width=\columnwidth]{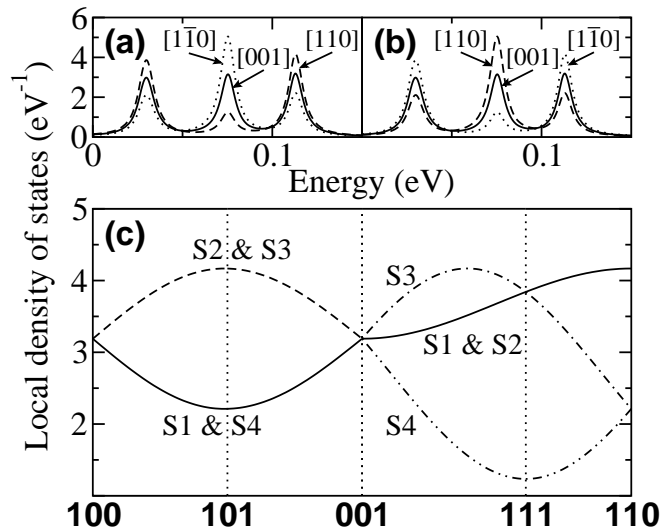}
\caption{ LDOS spectra at the sites (a) S1 or S2, and at the sites (b)
S3 or S4 as indicated in Fig.~\ref{fig:Mn3D}.  The solid, dotted and
dashed lines correspond to Mn spin orientations along $[001]$,
$[1\bar{1}0]$ and $[110]$ respectively. Zero energy is the valence
band maximum.  The energy broadening factor for the LDOS is
$5$~meV. (c) The spectral weight at $113$ meV as a function of Mn
spin orientation for the four As sites. }
\label{fig:sp}
\end{figure}

Although the energy spectrum (and thus the DOS) of the magnetic dopant
in a zincblende or diamond symmetry semiconductor is independent of
dopant spin orientation, the LDOS can vary considerably. The symmetry
of the LDOS would be tetrahedral ($T_d$) if there were no spin-orbit
interaction and also no Jahn-Teller distortion. For a Mn dopant
substituting for a Ga atom no Jahn-Teller distortion is
seen.\cite{Schneider1987} The actual symmetry of the LDOS for GaAs:Mn,
however, is lowered by the spin-orbit interaction, and the resulting
symmetry depends on the Mn spin orientation as shown in
Fig.~\ref{fig:Mn3D}. The LDOS is calculated as the imaginary part of
the Green's function, LODS$(\vR)=-\pi^{-1} {\rm Im }\left[{\rm tr}\,
G^R(\vR,\vR)\right]$.

To understand how the spatial symmetry and spin orientation are
correlated we consider the angular momentum character of the acceptor
states. The acceptor states found in our model are spin-polarized
antiparallel to the Mn $3d^5$ core spin due to the exchange coupling
between the GaAs valence states and Mn $d$ states.\cite{Vogl1985} Only
one acceptor state can be occupied by a hole, as the Coulomb
interaction prevents two holes from binding to the Mn dopant.  The
degeneracy of the singly occupied acceptor states with different
orbital angular momentum is lifted by the spin-orbit interaction, and
the lowest energy configuration has an orbital angular momentum
aligned antiparallel to the Mn core spin, leading to a composite $J=1$ spin associated with the Mn dopant.  As an $L_z\ne 0$ state of
an atom centered at ${\bf r}=0$ has a nodal line along $x=y=0$, the
probability density of the acceptor state should be small along the
line extending from the dopant parallel to the core spin
direction. Thus a contour surface of the LDOS at the lowest acceptor
level (shown in Fig.~\ref{fig:Mn3D}) has an approximately oblate shape
with the short axis aligned with the Mn core spin. The shape of the
level does not depend on which way the Mn core spin is pointing along
the axis --- only on the orientation of the axis itself.  One
consequence of this shape is anisotropic spin-spin
interaction,\cite{Schilfgaarde2001,Zarand2002,Brey2003,Fiete2003,Fiete2005,JMT2004,Mahadevan2004b,Timm2005}
in which the overlap is larger when the two Mn spins are perpendicular
to the axis that joins them. Figure \ref{fig:Mn3D} shows three
examples for which the Mn spin is oriented along one of the high
symmetry axes of a cubic crystal. When the Mn spin is aligned with the
(b) $[001]$, (c) $[1\bar{1}0]$ or (d) $[111]$ axis, the symmetry of
the state is lowered from $T_d$ to (b) $D_{2d}$, (c) $C_{2v}$ and (d)
$C_{3v}$. Although the spin direction is indicated with an arrow, the
panels in Fig.~\ref{fig:Mn3D} would look the same if the spin
direction is reversed, so for example the acceptor state would appear
as Fig.~\ref{fig:Mn3D}(b) for the Mn spin pointing in the $[001]$ or
$[00\overline{1}]$ direction.

Figure~\ref{fig:sp} shows the energy spectrum near the valence band
edge and the LDOS spectra at the four nearest-neighbor As sites
S1--S4. These sites are indicated by their labels in
Fig.~\ref{fig:Mn3D}(a). For a Mn at the origin, S1 is at
$(\frac{a}{4},-\frac{a}{4},\frac{a}{4})$, S2 at
$(-\frac{a}{4},\frac{a}{4},\frac{a}{4})$, S3 at
$(\frac{a}{4},\frac{a}{4},-\frac{a}{4})$ and S4 at
$(-\frac{a}{4},-\frac{a}{4},-\frac{a}{4})$, where $a$ is the lattice
constant.  As pointed out above, for all orientations of the Mn spin
the energies of the peaks in the spectra are the same. The Mn dopant
itself sits at a high symmetry location, and thus the LDOS at the Mn
site is independent of the spin orientation. The spectral weight of
the acceptor state at the four nearest-neighbor As sites evolves as
the Mn spin rotates. Results are shown in Fig.~\ref{fig:sp}(c) for S1
(solid line), S2 (dashed line), S3 (dot-dashed line) and S4
(double-dot-dashed line).  From $[100]$ to $[001]$ the spectra at S3
(S4) is the same as S2 (S1).  From $[001]$ to $[110]$ the spectra at
S2 is the same as S1. The orbital angular momentum aligns with the Mn
spin, and thus the spectral weight tends to increase when the site is
away from the spatial line drawn through the Mn dopant and parallel to
the Mn spin. Conversely the spectral weight tends to decrease when the
site is close to this spatial line. The variation among the
nearest-neighbor As sites is greater than a factor of two.  For sites
farther from the Mn dopant, the dependence of the LDOS on Mn spin
orientation can be much greater. The variability of the LDOS of the
continuum valence band states is considerably less; the LDOS for all
three spin orientations shown in Fig.~\ref{fig:sp} at the valence
maximum varies by $\sim 40\%$, but is the same within a percent below
the split-off energy ($-350$~meV).

\begin{figure}
\includegraphics[width=\columnwidth]{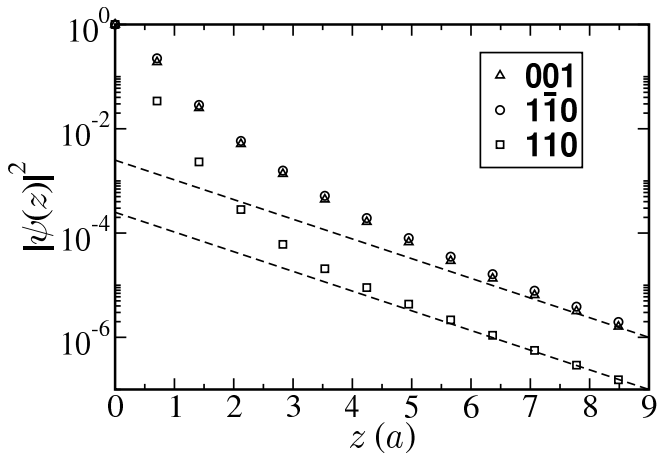}
\caption{ The probability density of the Mn acceptor state at the Ga
  sites along the $[110]$ direction ($z$ is the distance from the Mn
  dopant in units of $a$). The probability density drops significantly
  when the Mn spin is oriented in the $[110]$ direction (square)
  rather than in the perpendicular directions, $[001]$ (triangle) and
  $[1\bar{1}0]$ (circle). The probability density is normalized to
  unity at the Mn site. The dashed lines show $\sim
  e^{-2z/\lambda}$. }
\label{fig:wf}
\end{figure}

To show the LDOS variability for different Mn spin orientations as a
function of distance, we plot in Fig.~\ref{fig:wf} the probability
density of the Mn acceptor state at the Ga sites along the $[110]$
direction, which is usually the surface normal of a cleaved GaAs
sample. The background LDOS has been subtracted off so that the
behavior at large distances is visible, and the probability density
has been normalized to unity at the Mn site. The background LDOS for
the $5$~meV energy linewidth is about $10^{-3}$~eV$^{-1}$. A clear
feature of Fig.~\ref{fig:wf} is that the probability density drops
significantly when the Mn spin is oriented parallel to $[110]$ rather
than perpendicular to $[110]$. Once the distance is larger than about
4 atomic layers ($8$~\AA{} from the Mn site), the probability density
changes by roughly an order of magnitude as the Mn spin orientation is
changed from parallel to perpendicular to $[110]$.  At large distances
from the Mn site we expect the wave function follows a simple
exponential tail with a decay length we estimate based on the
effective masses of the heavy and light holes and on the acceptor
level energy. In our tight-binding model, the effective masses along
$[110]$ for heavy and light holes are about $m_{\rm hh}=0.5m_0$ and
$m_{\rm lh}=0.12m_0$.  For the acceptor level, the binding energy $E =
113$~meV and the decay length $\lambda=\hbar/(2m_{\rm
eff}E)^{(1/2)}\sim 13$~\AA{}, where the averaged effective mass is
$m_{\rm eff}=2m_{\rm hh}m_{\rm lh}/(m_{\rm hh}+m_{\rm lh})$. The
visible LDOS variability will be significantly reduced at distant
sites due to the presence of a background LDOS from the valence band
tail.

\begin{figure}
\includegraphics[width=\columnwidth]{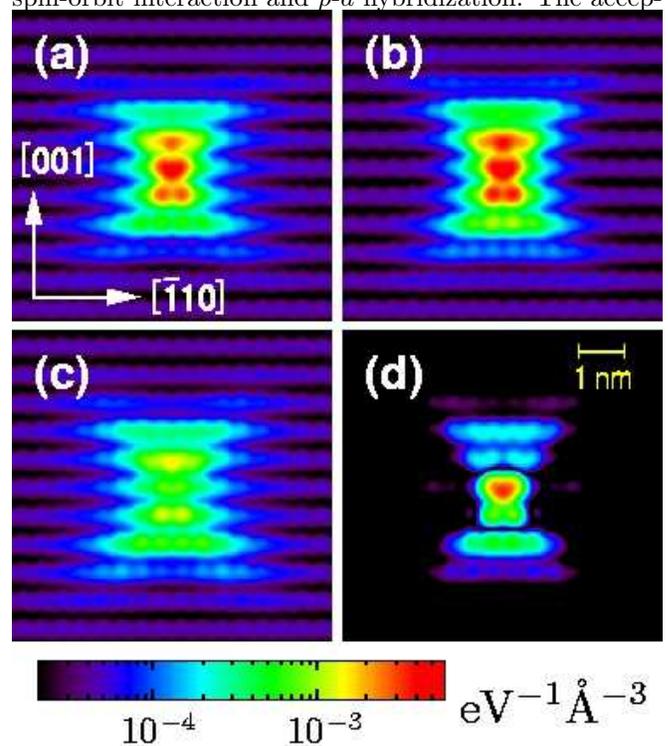}
\caption{ Cross-sectional view of the LDOS on the $(110)$ plane when
  the Mn dopant is four atomic layers (corresponding to $z=2^{(1/2)}a$
  in Fig.~\ref{fig:wf}) below the viewing layer. The Mn spin is
  orientated along (a) $[001]$ (b) $[1\bar{1}0]$ or (c) $[110]$.  The
  absolute difference between (b)+(c)$-$(a) and (a) is shown in
  (d). The shape of the apparent feature is largely the same for
  (a-c), but the amplitude differs considerably: the spectral weight
  at the center changes by $\sim 90$\% between (b) and (c), and by
  $\sim 15$\% between (a) and (b). The LDOS at each atomic site is
  spatially distributed according to a normalized Gaussian with a
  $2$\AA{} width. }
\label{fig:comp4}
\end{figure}

Figure \ref{fig:comp4} shows the field of view at the $(110)$ plane
four atomic layers from the Mn layer. This is the image most relevant
to detecting the ESR of a single Mn dopant in
GaAs. The procedure would be to cleave the GaAs along the $(110)$
plane, and locate an isolated Mn a few layers down from the surface
(here we have chosen four layers, but as shown in Fig.~\ref{fig:wf}
any layer would do except the surface layer). At the central Ga site
on the surface plane, here located $8$\AA{} from the Mn site in the
material below, there is approximately a $90$\% change in LDOS when
the Mn spin switches from parallel to the surface normal to
perpendicular to the surface normal.  The LDOS also differs for the
two Mn orientations perpendicular to the surface normal, although that
difference is considerably less ($15$\%). The overall shape of the
anisotropic acceptor state is very similar for all three spin
orientations chosen for Fig.~\ref{fig:comp4}(a-c), as the gross shape
is governed by the cubic symmetry of the crystal,\cite{Yakunin2004b}
but the degree of anisotropy (appearing as the magnitude of the
state's LDOS along this spatial slice) varies by up to $90$\%. These
characteristics suggest that atomic-scale resolution is not necessary
to distinguish the Mn spin axis from measurements such as those
suggested by Fig.~\ref{fig:comp4}.

The differences in LDOS shown in Fig.~\ref{fig:comp4} suggest that a
repetitive preparation of the Mn dopant in a particular spin state
(using ESR techniques) and measurement of the spin orientation at a
time delay (using the nonmagnetic STM measurement) would be able to
produce an oscillatory LDOS. The temporal frequency of the LDOS
created by the Mn spin's precession would be determined by the
magnetic field strength and $g$ factor of the Mn dopant
(measured\cite{Schneider1987} to be $2.77$), and the spatial structure
could be determined from Fig.~\ref{fig:comp4}. Based on the linewidth
in bulk ESR measurements done at $9.4$ GHz\cite{Schneider1987} ($<0.05$
Tesla), the spin coherence time of a single Mn spin is estimated to be
longer than $0.5$ ns. Ultrafast STM detection at $50$ GHz has been
achieved.\cite{Steeves1998} If the static magnetic field of an ESR
apparatus were directed parallel to $[001]$, then precession would
involve an oscillation between Fig.~\ref{fig:comp4}(b) and
Fig.~\ref{fig:comp4}(c). As the Mn spin has $J=1$, pulses of
oscillating magnetic fields could be used to manipulate the
populations of the $J_z=1$, $0$, $-1$ eigenstates. For
$J_{[001]}=\pm 1$ the LDOS would appear as Fig.~\ref{fig:comp4}(a). The
orbital wavefunctions transform according to $T_2$ in the tetrahedral
crystal, so the spatial structure of $|\psi_{J_z=0}|^2$ can be written
as $|\psi_{J_x=1}|^2+|\psi_{J_y=1}|^2-|\psi_{J_z=1}|^2$. Thus the
spatial structure of the $J_{[001]}=0$ state would be
Fig.~\ref{fig:comp4}(a) subtracted from the sum of
Fig.~\ref{fig:comp4}(b) and Fig.~\ref{fig:comp4}(c). The difference
between this $J_{[001]}=0$ state and the $J_{[001]}=\pm 1$ states is shown
in Fig.~\ref{fig:comp4}(d).

We have shown how the anisotropic spatial shape of the Mn acceptor
state, which principally originates from the cubic symmetry of the
GaAs lattice, is influenced by spin-orbit interaction. Calculations of
the Mn acceptor state were performed within a tight-binding model
including spin-orbit interaction and $p$-$d$ hybridization.  The
acceptor state has an approximately oblate shape with the short axis
aligned with the Mn core spin. The difference in the LDOS for Mn spins
oriented in two different directions at a site located $8$\AA{} away
from the Mn dopant can be as high as $90$\%. We suggest that the high
visibility of the spin orientation in the LDOS for Mn spins can be
probed by scanning tunneling microscopy, and this can lead to a
nonmagnetic detection scheme for single-spin ESR.

We acknowledge conversations with P. M. Koenraad, J. Levy,
A. Yu. Silov, and A. M. Yakunin.  This work was supported by the USARO
under MURI Grant No. DAAD19-01-1-0541.

\end{document}